\documentstyle[epsfig,12pt]{article}

\begin{document}
\begin{titlepage}
\begin{flushright}
UFIFT-HEP-01-5 \\ Revised
\end{flushright}
\vspace{.4cm}
\begin{center}
\textbf{Homogeneous and Isotropic Charge Densities \\
to Accelerate the Universe}
\end{center}
\begin{center}
M. Brisudova$^{\dagger}$ and R. P. Woodard$^{\ddagger}$
\end{center}
\begin{center}
\textit{Department of Physics \\ University of Florida \\
Gainesville, FL 32611}
\end{center}
\begin{center}
W. H. Kinney$^*$
\end{center}
\begin{center}
\textit{Institute for Strings, Cosmology and Astroparticle Physics \\
Columbia University \\ New York, NY 10027}
\end{center}
\begin{center}
ABSTRACT
\end{center}
\hspace*{.5cm} Various theoretical obstacles are associated with a 
homogeneous and isotropic distribution of ``charge'' which is subject 
to a repulsive, long-range force. We show how these can be overcome, 
for all practical purposes, by the simple device of endowing the 
particle which carries the force with a small mass. The resulting 
situation may be relevant to a phase of cosmological acceleration 
which is triggered by the approach to masslessness of such a force 
carrier.
\begin{flushleft}
\end{flushleft}
\vspace{.4cm}
\begin{flushleft}
$^{\dagger}$ {\it On leave from Physics Inst., SAS, Bratislava, Slovakia. }
e-mail: brisuda@phys.ufl.edu \\
$^{\ddagger}$ e-mail: woodard@phys.ufl.edu \\
$^*$ email: kinney@physics.columbia.edu
\end{flushleft}
\end{titlepage}

Recent observations of Type Ia supernovae with high redshift indicate that the
universe is entering a phase of cosmological acceleration \cite{reiss,perlm}.
Identifying the causative agent is both a challenge and an opportunity for
fundamental theory. It could be a cosmological constant, the need for which
was suggested on the basis of other evidence even before the supernovae
results \cite{KT,OS,LLVW}. Scalars will also work because one can construct 
a potential to support {\it any} homogeneous and isotropic geometry 
for which the Hubble constant does not increase.\footnote{For the 
construction see section 2 of \cite{TW}.} Minimally coupled scalars becoming
dominant at late times was also suggested before the supernovae results
\cite{PR,Wett,WCL,FJ}. Since then such models have been dubbed, 
``quintessence'' \cite{CDS} and have received extensive study
\cite{CLW,ZWS,BM,BCN,Arm}. Nonminimal couplings have also been explored
\cite{APMS} and recent inspiration has been derived from string theory 
\cite{FKPMP,HKS} and extra dimensions \cite{Maeda,HL}. It has even been
suggested that quantum effects may be responsible \cite{LPR}.

In the absence of compelling observational or theoretical support for any of
the existing scenarios it is worth considering what else might be driving 
the late time acceleration we seem to be seeing. Since different portions 
of an accelerating universe seem to be pushing one another apart an obvious 
alternative is that this may actually be the case. That is, suppose some 
constituent of the current energy density carries a charge --- for example, 
baryon number --- which couples to a repulsive force that only became 
long-range late in evolution.

Powerful objections seem to preclude the realization of this scenario with\-in
the context of a homogeneous and isotropic cosmology. If the charge density is
homogeneous then the total charge must be the 3-volume times this density. On a
closed 3-manifold the total charge of any infinite range force field must be
zero, so the density would have to vanish. The general phenomenon is known as 
a {\it linearization instability} \cite{Wald}. One can impose a nonzero,
homogeneous charge density on an open manifold, but not without breaking
isotropy through the selection of a direction for the lines of force. Further,
the charge density could only be instantaneously homogeneous because different
regions would necessarily feel different force fields.

These objections can be made more concrete within the context of scalar
electrodynamics, the Lagrangian for which is,
\begin{equation}
{\cal L} = -\frac14 F_{\alpha\beta} F_{\rho\sigma} g^{\alpha\rho} g^{\beta
\sigma} \sqrt{-g} + (D_{\mu} \phi)^* (D_{\nu} \phi) g^{\mu\nu} \sqrt{-g} -
V(\phi^* \phi) \sqrt{-g} \; ,
\end{equation}
where the covariant derivative is $D_{\mu} \equiv \partial_{\mu} -i e A_{\mu}$.
The Euler-Lagrange equations for the scalar and vector potential are,
\begin{eqnarray}
{\delta S \over \delta \phi^*} & = & -D_{\mu} \left(\sqrt{-g} g^{\mu\nu}
D_{\nu} \phi\right) - \phi V'(\phi^* \phi) \sqrt{-g}  = 0 \; , \\
{\delta S \over \delta A_{\mu}} & = & \partial_{\nu} \left(\sqrt{-g} g^{\nu
\rho} g^{\mu \sigma} F_{\rho\sigma}\right) + i e \sqrt{-g} g^{\mu\nu}
\left(\phi^* D_{\nu} \phi - \phi D^*_{\nu} \phi^*\right) = 0 \; . \label{max}
\end{eqnarray}
The two sources for this system are the current density,
\begin{equation}
J_{\mu} \equiv i e \left(\phi D^*_{\mu} \phi^* -\phi^* D_{\mu} \phi\right) \; ,
\label{J}
\end{equation}
and the stress-energy tensor,
\begin{eqnarray}
\lefteqn{T_{\mu\nu} \equiv {2 \over \sqrt{-g}} {\delta S \over \delta
g^{\mu\nu}} = -F_{\mu\rho} F_{\nu\sigma} g^{\rho\sigma} + \frac14 g_{\mu\nu}
F_{\alpha\beta} F_{\rho\sigma} g^{\alpha\rho} g^{\beta\sigma}} \nonumber \\
& & \qquad \qquad + 2 (D_{\mu} \phi)^* D_{\nu} \phi - g_{\mu\nu}
\left(g^{\rho\sigma} (D_{\rho} \phi)^* D_{\sigma} \phi - V(\phi^*\phi)\right)
\; .
\end{eqnarray}

Assuming spatial flatness in addition to homogeneity and isotropy, the metric
can be put in the form,
\begin{equation}
g_{\mu\nu} dx^{\mu} dx^{\nu} = dt^2 - a^2(t) d\vec{x} \cdot d\vec{x} \; .
\end{equation}
The scalar can be written, $\phi(t) = f(t) e^{i \theta(t)}$, in terms of its
magnitude $f(t)$ and its phase $\theta(t)$. The only nonzero vector potential
can be $A_0(t)$ and (\ref{max}) can be solved for it uniquely to give,
\begin{equation}
A_0(t) = {i e \over 2 e^2 \phi^* \phi} \left(\phi \partial_0 \phi^* - \phi^*
\partial_0 \phi\right) = {\dot{\theta}(t) \over e} \; .
\end{equation}
This seems to be representing a self-interaction of nonzero charge density but one
must bear in mind that the scalar current density (\ref{J}) involves the vector
potential. From the scalar's covariant time derivative,
\begin{equation}
D_0 \phi = \dot{f}(t) e^{i\theta(t)} \; ,
\end{equation}
we see that the actual charge density vanishes, while the stress tensor depends
only upon the scalar magnitude $f(t)$.

Of course the frustrating result we have just found merely confirms the general
objections: the only charge density consistent with homogeneity and isotropy is
zero for an infinite range force. One cannot evade this fact mathematically,
but it {\it can} be circumvented for practical purposes by the simple device of
giving the force a finite range which can still be much larger than the Hubble
radius. The problem really derives from Gauss's Law, and it can be understood
in its simplest form on the flat, $1+1$ dimensional manifold $R^1 \times S^1$.
The equation for the Coulomb Green's function is,
\begin{equation}
-{d^2 \over dx^2} G(x,x') = \delta(x-x') \; .
\end{equation}
There is no solution with $x= \pm L$ identified, so one cannot have a nonzero
total charge on $S^1$ or, it turns out, on any closed spatial manifold.

The usual procedure is to subtract the zero mode and solve for the restricted
Green's function appropriate to a charge density with zero total charge. The
relevant equation is,
\begin{equation}
-{d^2 \over dx^2} G_r(x,x') = \delta(x-x') - {1 \over 2 L} \; ,
\end{equation}
and, up to a constant, the solution is,
\begin{equation}
G_r(x,x') = {(L+x-x')^2 \over 4 L} \theta(x'-x) + {(L-x+x')^2 \over 4L}
\theta(x-x') \; .
\end{equation}
What we are advocating instead is to add a small mass to attain the equation,
\begin{equation}
\left\{ -{d^2 \over dx^2} + m^2 \right\} G_m(x,x') = \delta(x-x') \; .
\end{equation}
The unique periodic solution is,
\begin{equation}
G_m(x,x') = {\cosh[m(L + x - x')] \over 2 m \sinh(mL)} \theta(x'-x) +
{\cosh[m (L - x + x')] \over 2 m \sinh(2m)} \theta(x-x') \; . \label{G_m}
\end{equation}
Note that the mass can be very much smaller than $1/L$; as long as $m \neq 0$
the equation can be solved. Further, the solution makes physical sense for
small $m$. Expanding $G_m(x,x')$ for small $m$ gives a constant term which
diverges like $1/m^2$, followed by $G_r(x,x')$ and terms which vanish with $m$.
It is this first term which is relevant for a homogeneous and isotropic
cosmology.

The preceding discussion suggests that we wish to give the vector a mass. This
entails breaking gauge invariance but one can at least preserve general
coordinate invariance with the Lagrangian,\footnote{In a model with two scalars
one could preserve gauge invariance by generating the photon mass through
spontaneous symmetry breaking.}
\begin{eqnarray}
\lefteqn{{\cal L} = -\frac14 F_{\alpha\beta} F_{\rho\sigma} g^{\alpha\rho}
g^{\beta\sigma} \sqrt{-g} + \frac12 m^2 A_{\mu} A_{\nu} g^{\mu\nu} \sqrt{-g}}
\nonumber \\
& & \qquad \qquad + (D_{\mu} \phi)^* (D_{\nu} \phi) g^{\mu\nu} \sqrt{-g} -
V(\phi^* \phi) \sqrt{-g} \; ,
\end{eqnarray}
Within the context of homogeneity and isotropy the unique solution for the
vector potential is,
\begin{equation}
A_0(t) = {ie \over m^2 + 2 e^2 \phi^* \phi} \left(\phi \partial_0 \phi^* -
\phi^* \partial_0 \phi\right) = {2 e f^2(t) \dot{\theta}(t) \over m^2 + 2 e^2
f^2(t)} \; .
\end{equation}
The scalar's covariant time derivative becomes,
\begin{equation}
e^{-i \theta} D_0 \phi = \dot{f} + {i m^2 f \dot{\theta} \over m^2 + 2 e^2 f^2}
\; ,
\end{equation}
and the charge density (\ref{J}) is,
\begin{equation}
J_0(t) = -m^2 A_0(t) = {-2 e m^2 f^2(t) \dot{\theta}(t) \over m^2 + 2 e^2
f^2(t)} \; . \label{J_01}
\end{equation}
Despite the breaking of gauge invariance, the scalar equations of motion still
imply that this charge density is conserved,
\begin{equation}
{d \over dt} \left( {\mbox{} \over \mbox{}} a^3(t) J_0(t) \right) = 0 \;
\end{equation}
If $n_0$ is the current number density of charges (when $a(t_0) = 1$) then
we can isolate the time dependence as follows,
\begin{equation}
J_0(t) = {e n_0 \over a^3(t)} \; . \label{J_02}
\end{equation}
Since general coordinate invariance was maintained the stress tensor is still
conserved.

In considering how things change with $m$ and $e$ it is useful to express the
phase using relations (\ref{J_01}) and (\ref{J_02}),
\begin{equation}
\dot{\theta}(t) = - \left({1 \over 2 f^2(t)} + {e^2 \over m^2}\right) {n_0
\over a^3(t)} \; .
\end{equation}
In these variables the photon and scalar kinetic terms are,
\begin{equation}
\frac12 m^2 A_0^2 + (D_0 \phi)^* D_0 \phi = \dot{f}^2 + {m^2 f^2 \dot{\theta}^2
\over m^2 + 2 e^2 f^2} = \dot{f}^2 + \left({1 \over 4 f^2} + {e^2 \over 2 m^2}
\right) \left({n_0 \over a^3}\right)^2
\end{equation}
It follows that the energy density and pressure are,
\begin{eqnarray}
\rho & = & \dot{f}^2 + \left({1 \over 4 f^2} + {e^2 \over 2 m^2} \right)
\left({n_0 \over a^3}\right)^2 + V(f^2) \; , \label{rho} \\
p & = & \dot{f}^2 + \left({1 \over 4 f^2} + {e^2 \over 2 m^2} \right)
\left({n_0 \over a^3}\right)^2 - V(f^2) \; . \label{p}
\end{eqnarray}
The extra term due to the vector interaction is the one proportional to $e^2/2
m^2$. Note that it makes physical sense. Turning on the interaction, with fixed
charge density per unit charge {\it raises} the energy, as one expects for a
repulsive interaction. Similarly, the energy density diverges like $1/m^2$ for
small $m$, just as the massive Coulomb Green's function (\ref{G_m}) does.

The obvious phenomenological application for this trick is to give a model in
which some constituent of the current energy density contains a uniformly
distributed charge coupled to a force field that has recently been driven
nearly massless. In this case $m^2$ would be the norm of some other complex
scalar --- call it $\psi$ --- whose phase is negligible. Suppose that the 
minimum of the total potential is at $\psi = 0$. Then as $\psi$ tries to roll
down to this minimum the interaction gives rise to a peculiar sort of {\it 
electromagnetic barrier} which pushes $\psi$ back up its potential. By reducing
the kinetic energy (which obeys $p_K = \rho_K$) and enhancing the potential 
energy (which obeys $p_P = -\rho_P$) this must favor cosmological acceleration.

For certain models the effect can be enough to make the time average 
deceleration parameter negative and in fact close to $-1$. What actually 
happens is that the field oscillates about the minimum of the effective 
potential obtained from adding the electromagnetic barrier (which is time 
dependent through the factor of $1/a^6$) to the original potential. Since the 
electromagnetic barrier is very steep and Hubble friction is low there, the
field recoils sharply off and spends most of its time at high potential where 
Hubble friction is greatest. Therefore the time average of the deceleration 
parameter is dominated by the period spent at high potential with low kinetic 
and electromagnetic contributions to the energy density. Detailed numerical 
simulations have been done and will be presented in a subsequent paper 
\cite{BKW}.

It is also interesting to note that the effective range of the force changes
instantaneously with the mass. The virtual vector quanta which carry the
repulsive interaction from one patch of charge density to another lose or
acquire mass in route. This has the curious consequence that, even though the
electromagnetic barrier forms due to an interaction becoming long range, one
does not have to wait for distant regions to come into causal contact with one
another after the range changes.

\vskip 1cm
\centerline{\bf Note Added}

After the completion and release of this work we learned of a paper treating
repulsive interactions in general terms from the context of phenomenological
models \cite{ZSBP}. It seems to us that our technique may provide a class of
Lagrangian field theories which explicitly realize these ideas. We also became 
aware of a somewhat related class of models which exploit the kinetic energy in
the phase of a complex scalar field \cite{GH,BCK}. The energy and pressure in 
these theories agree with (\ref{rho}-\ref{p}) for $e=0$. Our models include a 
long range repulsive potential, in addition to the kinetic energy of the phase.
This allows a wider range of possibilities in which the mass of the force 
carrier is generated by another complex scalar, or by some dynamical mechanism.
It may also be relevant to the tendency for these models to decay into Q-balls
\cite{Kasuya}. For note that the repulsive potential must survive, and must
continue to push the universe apart, even if the charge is been bundled into 
Q-balls.

\vskip 1cm
\centerline{\bf Acknowledgments}

We thank M. Axenides and L. Perivolaropoulos for acquainting us with the
simultaneous work on spintessence. This work was partially supported by DOE 
contract DE-FG02-97ER\-41029 and by the Institute for Fundamental Theory.

\end{document}